\documentstyle[12pt]{article}
\begin{document}
\baselineskip=25pt   
\def\half{{\displaystyle {1 \over 2}}}
\def\thalf{{\textstyle {1 \over 2}}}
\def\<{\langle}
\def\>{\rangle}
\def\pdiv#1{{\partial \over {\partial #1}}}

\setcounter{secnumdepth}{1}

\pagestyle{plain}
\begin{center}
{\Large \bf Scaling in high-temperature superconductors}

by

Ian D Lawrie$^{\dag}$

Department of Physics, The University of Leeds, Leeds LS2 9JT, UK

(March 1994)

\end{center}

\begin{center}
{\bf Abstract}
\end{center}

A Hartree approximation is used to study the interplay of two kinds of
scaling which arise in high-temperature superconductors, namely
critical-point scaling and that due to the confinement of electron
pairs to their lowest Landau level in the presence of an applied
magnetic field.  In the neighbourhood of the zero-field critical
point, thermodynamic functions scale with the scaling variable
$(T-T_{c2}(B))/B^{1/2\nu}$, which differs from the variable $(T -
T_c(0))/B^{1/2\nu}$ suggested by the gaussian approximation.
Lowest-Landau-level (LLL) scaling occurs in a region of high field
surrounding the upper critical field line but not in the vicinity of
the zero-field transition.  For YBaCuO in particular, a field of at
least 10 T is needed to observe LLL scaling.  These results are
consistent with a range of recent experimental measurements of the
magnetization, transport properties and, especially, the specific heat
of high-$T_c$ materials.
\vfill \noindent

\noindent $^{\dag}$e-mail address:  i.d.lawrie@uk.ac.leeds
\newpage
\noindent{\bf 1.\ Introduction}

The superconducting transition in conventional low-$T_c$ materials is
well described by the Ginzburg-Landau mean field theory.  Principally
because of the large correlation volume in these materials, the region
in which critical fluctuations might be important is too small to be
accessible experimentally.  In high-temperature superconductors, by
contrast, the critical region may be much larger.  Widely varying
theoretical estimates of the size of this region are obtained
according to details of the criterion employed$^1$, but marked
deviations from mean-field behaviour have been observed$^{2 - 6}$ over
a temperature range of the order of 10 K above and below $T_c$.

Theoretical expectations of the kind of critical behaviour which might
be observed are somewhat confused.  If fluctuations in the magnetic
vector potential can be ignored, then the zero-field transition ought
to be a critical point in the universality class of the 3-dimensional
XY model, as is the superfluid transition in $^4$He.  When magnetic
fluctuations are included, a renormalization-group analysis by
Halperin, Lubensky and Ma$^7$ for a $(4-\epsilon)$-dimensional system
reveals a runaway of renormalization-group trajectories, which these
authors interpreted as a signal of a weakly first-order transition.
This interpretation is confirmed by an explicit construction of the
free energy$^8$.  On the other hand, a renormalization-group analysis
in $(2 + \epsilon)$ dimensions$^9$ indicates a second-order
transition in the universality class of the $CP^{N-1}$ model in the
limit $N \rightarrow 1$, while a lattice simulation of Dasgupta and
Halperin$^{10}$ is consistent with inverted XY critical behaviour. In
high-$T_c$ cuprates, the region in which a first-order transition or
inverted XY behaviour might be detected is probably extremely small.
These are strongly type-II materials with penetration depths in excess
of 1000 \AA. In this situation, renormalization-group trajectories
pass very close to the ordinary XY fixed point, suggesting that
magnetic fluctuations can indeed be ignored except in a very narrow
range of temperatures near $T_c$.

Experience of critical phenomena in, for example, fluids and magnets
suggests that, in the presence of an applied magnetic field $B$, there
should be a critical region in which thermodynamic quantities assume
the scaling form $A(T,B) = B^{\alpha_A}{\cal A}(x)$, where $\alpha_A$
is a critical exponent associated with the quantity $A$ and the
scaling variable is an appropriate ratio of scaling fields.  In
principle, the correct scaling fields would emerge from a
renormalization-group analysis, but this analysis is very difficult in
the presence of an applied field.  An early calculation of
Prange$^{11}$ using the gaussian approximation suggests that $B$
occurs in the combination $B\xi^2$, where $\xi$ is the zero-field
coherence length, so that the scaling variable should be $x =  (T-T_c)
/B^{1/2\nu}$, where $\nu$ is the coherence-length exponent.  We shall
argue that this is not quite correct, however, and that the scaling
variable should be $x = (T - T_{c2}(B))/B^{1/2\nu}$, where the line $T
= T_{c2}(B)$ is a renormalized version of the line usually denoted by
$H_{c2}(T)$, the upper critical field in mean field theory. In the
gaussian approximation, with $\nu = \half$, these two scaling
variables differ only by an additive constant.

A different scaling form for thermodynamic functions (which is not
directly associated with a phase transition) arises in the lowest
Landau level (LLL) approxima\-tion$^{12 - 14}$, which is usually
thought to be valid in the neighbourhood of the $H_{c2}(T)$ line.
Here, the appropriate scaling variable is $y = (T -
T_{c2}(B))/B^{\phi}$, where the exponent $\phi$ has the value $\phi =
{2 \over 3}$ in 3 dimensions or $\phi = \thalf$ in 2 dimensions. This
scaling behaviour is well verified experimentally for conventional
superconductors$^{12 - 16}$, but the type of scaling which applies to
high-$T_c$ materials is at present a matter of some controversy. It
has been claimed by Welp {\it et al}$^{\, 17}$ that magnetization,
conductivity and specific heat data for YBaCuO are consistent with LLL
scaling. On the other hand, Inderhees {\it et al}$^{\, 4}$ and Salamon
{\it et al}$^{\, 5}$ claim that experimental data is more nearly
consistent with critical-point scaling. In fact, the predicted scaling
forms for the magnetization and conductivity are probably too similar
to be distinguished experimentally. For the specific heat, the LLL
scaling fit exhibited by Welp {\it et al} is rather poor, and is
achieved only by the introduction of a prefactor which has no
theoretical basis.

In this work, we use a Hartree approximation to study the interplay of
critical-point and LLL scaling in an isotropic, $d$-dimensional
system.  Although materials such as YBaCuO are anisotropic, layered
systems, we expect that the results of this study should provide a
reasonable qualitative guide to the scaling behaviour in the vicinity
of the critical point, where the coherence length is much larger than
the interlayer spacing.  The Hartree approximation is in any case too
crude to give accurate numerical estimates either of critical
exponents or of scaling functions. We find that critical scaling may
be expected in a region of small fields and temperatures near
$T_c(0)$, while LLL scaling occurs in a region which surrounds the
$H_{c2}(T)$ line, but stops short of the zero field critical point at
a field value which we estimate at between 10 and 100 T. The Hartree
approximation is described in section 2 below, and a criterion for the
validity of the LLL approximation is obtained in section 3.  Sections
4 - 6 discuss scaling behaviour of the field-dependent coherence
length, the specific heat and the electrical conductivity.  A
comparison with recent experimental measurements is made in section 7,
and our conclusions are summarized in section 8.

\noindent{\bf 2.\ Hartree approximation}

The Ginzburg-Landau-Wilson reduced Hamiltonian may be written in a
standard form as
$$
{\cal H} = \int d^dx\left[\vert({\bf \nabla} - ie\hbox{\bf A})
\phi\vert^2 + t_0\vert\phi\vert^2 +
\half\lambda\vert\phi\vert^4\right]                         \eqno(2.1)
$$
where, in the critical region, $t_0$ can be taken as linear in
temperature and $\lambda$ as a constant. We assume throughout that the
magnetic flux density ${\bf B} = \nabla \times {\bf A}$ is uniform,
and equal to the applied field. Thus, the vector potential $A$ is not
a fluctuating variable, and the expectation value of a quantity
$f(\phi)$ is
$$
\< f(\phi)\> = \int {\cal D}\phi f(\phi)\exp(-{\cal H})/\int {\cal
D}\phi\exp(-{\cal H})\ .                                    \eqno(2.2)
$$
For a $d$-dimensional system, a convenient gauge choice is ${\bf A} =
\thalf B(-y, x, {\bf 0})$, where ${\bf 0}$ denotes the components in
the (d - 2) dimensions ``parallel'' to the magnetic field.
We implement an approximation of the Hartree type by introducing an
approximate Hamiltonian
$$
{\cal H}_0  = \int d^dx\left[\vert({\bf \nabla} - ie\hbox{\bf A})
\phi\vert^2 + (t_0 + \mu)\vert\phi\vert^2\right]\ ,         \eqno(2.3)
$$
where $\mu$ is determined self-consistently by requiring that ${\cal
H}$ and ${\cal H}_0$ have the same expectation value in the ensemble
of ${\cal H}_0$:
$$
\<{\cal H} - {\cal H}_0\>_0 = {\lambda \over 2}\<\vert\phi\vert^4\>_0
- \mu\<\vert\phi\vert^2\>_0 = 0\ .                          \eqno(2.4)
$$
The expectation values are easily expressed in terms of a sum over
Landau levels as $\<\vert\phi\vert^2\>_0 = I$ and
$\<\vert\phi\vert^4\>_0 = 2I^2$, where
$$
I = {{2eB}\over{4\pi}}\sum_n\int{{d^{d-2}k}\over{(2\pi)^{d-2}}}{1
\over{k^2 + 2eBn + t_0 + \mu +eB}}\ .                       \eqno(2.5)
$$
On defining
$$
\tilde{t} = t_0 +eB + \mu\ ,                               \eqno (2.6)
$$
we obtain the constraint equation (2.4) in the form
$$
\tilde{t} = t_0 + eB + {\lambda \over{4\pi}}2eB \sum_n\int{{d^{d-2}k}
\over{(2\pi)^{d-2}}}{1 \over{k^2 + 2eBn + \tilde{t}}}\ .    \eqno(2.7)
$$

The sum and integral in this expression is divergent, unless the
integration is restricted by an upper cutoff, of the order of an
inverse lattice spacing.  However, this divergence can be eliminated
by an additive renormalization of the temperature.  On carrying out
the sum and angular integrations, we find
$$
\tilde{t} = t + eB + \hat{\lambda}(2eB)^{d/2-1}f(\tilde{t}/2eB)\ ,
                                                            \eqno(2.8)
$$
where $\hat{\lambda} = \lambda/(4\pi)^{d/2}$, the renormalized
temperature variable is $t = t_0 + \hat{\lambda}\int_0^\infty
dx\,x^{-d/2}$ and
$$
f(z) = \int_0^\infty dx\,x^{-d/2}\left[{{xe^{-xz}}\over{1-e^{-x}}}-1
\right]\ .                                                  \eqno(2.9)
$$
For $2 < d < 4$, this integral is finite.  We are not able to evaluate
it analytically, but its limiting behaviour for small and large values
of $z = \tilde{t}/2eB$ can be obtained straightforwardly. When z is
small, we have
$$
f(z)\approx f_0z^{d/2-2}\ ,                                \eqno(2.10)
$$
where $f_0 = \int_0^\infty dx\,x^{1-d/2}e^{-x}$ ($=\surd\pi$ for
$d=3$). In this limit, the constraint equation reads
$$
\tilde{t} \approx t + eB + \hat{\lambda}f_0(2eB)\tilde{t}^{d/2 - 2}\ .
                                                           \eqno(2.11)
$$
This limit corresponds to the lowest Landau level (LLL) approximation,
where the sum in (2.7) is approximated by the term $n = 0$. In this
approximation, $\tilde{t}$ can be expressed in a scaling form, which
is inherited by various thermodynamic functions$^{12}$, namely
$$
\tilde{t} \approx (\hat{\lambda}f_0eB)^{\phi}\tau_{LLL}(y)\ ,
                                                           \eqno(2.12)
$$
where $\phi = 2/(6-d)$ (so $\phi = 2/3$ in 3 dimensions), $y = (t +
eB)/(\hat{\lambda}f_0eB)^{\phi}$ and the scaling function $\tau_{LLL}$
is the solution of
$$
\tau_{LLL}(y) = y + 2\tau_{LLL}(y)^{1 - 1/\phi}\ .         \eqno(2.13)
$$
In the opposite limit, $z \rightarrow \infty$, corresponding to very
low fields, we have
$$
f(z) \approx -f_\infty z^{d/2-1}\ ,                        \eqno(2.14)
$$
where $f_\infty = \int_0^\infty dx\,x^{-d/2}(1 - e^{-x})$ ($=2\surd\pi
$ for $d=3$).

\noindent{\bf 3.\ Validity of the lowest Landau level approximation}

One can, of course, make the approximation of neglecting all Landau
levels except the lowest without also invoking the Hartree
approximation, and perturbative calculations based on such an
approximation have been pursued by several authors$^{14, 15, 18 }$.
One finds that the scaling form (2.12) persists, though the scaling
function is no longer that given by solving (2.13). It is clearly
essential to know where in the phase diagram the LLL approximation is
likely to be valid, and we address this question within the Hartree
approximation in the following way. We suppose that the function
$f(z)$ is well approximated by (2.10) whenever $z$ is smaller than
some fiducial value, say $\epsilon$. For $z = \tilde{t}/2eB =
\epsilon$, the constraint (2.11) reads
$$
t = -eB[1-2\epsilon + 2 \hat{\lambda}f_0(2eB\epsilon)^{-\psi}]\ ,
                                                            \eqno(3.1)
$$
where $\psi = \phi^{-1}-1 = 2 - d/2$ ($\psi =1/2$ in three dimensions)
and defines a locus in the ($t$,$B$) plane, which is shown
schematically in figure 1. Points for which $z < \epsilon$ lie below
this line and this, therefore, is the region in which we might expect
the LLL approximation to be valid.  However, our analysis is based on
the assumption that the order parameter $\<\phi\>$ is negligibly
small.  This assumption presumably becomes invalid at some distance
below the mean-field $H_{c2}(T)$ line $t = -eB$, but we are unable
to determine whether some form of the LLL approximation survives with
$\<\phi\> \ne 0$.  Supposing that the order parameter is indeed
negligible above the line labelled $\<\phi\>\sim 0$ in figure 1 (whose
location we cannot determine precisely), we are led to the
conservative conclusion that the LLL approximation should be valid in
the roughly wedge-shaped region labelled LLL in the figure. This
region encloses a high-field, low-temperature portion of the line $t =
-eB$, but stops short of the critical point $t = B = 0$. As the
condition for the accuracy of the LLL approximation is made more
stringent, by reducing the value of $\epsilon$, the wedge recedes to
higher fields and lower temperatures.

\noindent{\bf 4.\ Scaling of the coherence length}

{}From the appearance of $\tilde{t}$ in the propagator in (2.7), we can
identify this quantity in terms of a temperature- and field-dependent
coherence length $\xi(t,B)$ as $\tilde{t} = \xi(t,B)^{-2}$. We find
from the constraint equation (2.8) that it can be expressed in a
two-parameter scaling form as
$$
\tilde{t} = (2eB)\tau_B(\theta, \delta)\ ,                  \eqno(4.1)
$$
where the scaling variables are
$$
\theta = \hat{\lambda}^{-1}(t + eB)(2eB)^{-1/2\nu}\ ,       \eqno(4.2)
$$
$$
\delta = \hat{\lambda}^{-1}(2eB)^{\omega/2}\ ,              \eqno(4.3)
$$
with exponents $\nu = 1/(d-2)$ and $\omega = 4-d$, and the scaling
function $\tau_B$ is the solution of
$$
\delta\tau_B = \theta + f(\tau_B)\ .                        \eqno(4.4)
$$
It is straightforward to show that $\tau_B$ has a double power series
expansion in $\theta$ and $\delta$. For small fields and temperatures
close to $T_c$, we can take the limit $\delta \rightarrow 0$ with
$\theta$ fixed, to obtain the one-parameter scaling form
$$
\tilde{t}\approx (2eB)\tau_B(\theta, 0) = (2eB)f^{-1}(-\theta)\ .
                                                            \eqno(4.5)
$$
This should be valid in a region near the critical point $t=B=0$,
indicated schematically by the shaded ``Critical'' region in figure 1.
According to (2.10) and (2.14), the critical-point scaling function
$\tau_B(\theta,0) = f^{-1}(-\theta)$ has the limiting forms
$$
\tau_B(\theta,0) \approx \left({\theta \over f_{\infty}}\right)^{2/(d
-2)}\ ,\ \ \theta \rightarrow + \infty \ ,                  \eqno(4.6)
$$
$$
\tau_B(\theta,0) \approx \left(- {\theta \over f_0}\right)^{-2/(4-d)}
\ ,\ \ \theta \rightarrow - \infty \ .                      \eqno(4.7)
$$

In general, the regions where the LLL scaling form (2.12) and the
critical scaling form (4.5) hold must be distinct. There may, however,
be a crossover region in which both forms are approximately valid.
This requires
$$
\tau_B(\theta, 0) \sim \theta^{-\sigma}                     \eqno(4.8)
$$
and
$$
\tau_{LLL}(y) \sim y^{-\sigma} \ ,                          \eqno(4.9)
$$
where $\sigma = 2\nu(1 - \phi)/(2\nu\phi - 1)$.  In the Hartree
approximation, we have $\sigma = 2/(4-d)$ and we find from (2.13) and
(4.7) that there is such a crossover region, namely the region of very
low field below $T_c$, where $\theta$ and $y$ are both large and
negative.  It is, however, in this region that the approximation
$\<\phi\> \approx 0$ is likely to fail, so (4.8) and (4.9) are not
necessarily meaningful in this region.  Moreover, we do not know
whether these limiting forms of the scaling functions are valid beyond
the Hartree approximation, so it is not clear whether the crossover
region would be accessible in real materials.

In the usual way, the scaling relation (4.1) can be reformulated as
$$
\tilde{t}=\vert t+eB\vert^{2\nu}\tau_t^{\pm}(\beta,\eta)\ ,\eqno(4.10)
$$
where $\beta = 2eB\vert t+eB\vert^{-2\nu}$, $\eta = \vert
t+eB\vert^{\omega\nu}$ and the $+$ and $-$ branches of the scaling
function refer to $t > -eB$ and $t < -eB$. It is straightforward to
show that $\tau_t^{\pm}$ has a double power series expansion in
$\beta$ and $\eta$. The form (4.1) is, however generally the more
convenient.

\noindent {\bf 5.\ Scaling of the specific heat}

Identification of the specific heat within the Hartree approximation
is somewhat ambiguous.  On the one hand, we can define a Hartree
approximation to the free energy density, $F_0 = - V^{-1}\ln
\left[\int{\cal D}\phi \exp(-{\cal H}_0)\right]$, where ${\cal H}_0$
is the approximate Hamiltonian introduced in (2.3), and identify $C =
\partial^2F_0/\partial t^2$. On the other hand, we can identify the
entropy density of the original Ginzburg-Landau-Wilson model (2.1) as
$S = -\thalf\<\vert\phi\vert^2\>$ and the specific heat as $C =
\partial S/\partial t$, and evaluate this quantity in the Hartree
approximation:
$$
C = -\half {\partial\over{\partial t}}\<\vert\phi\vert^2\>_0\ .
                                                            \eqno(5.1)
$$
These two definitions are not equivalent. The latter definition was
adopted, for example, by Bray$^{12}$ and is the one we use here. It is
slightly simpler, and agrees with the natural definition of the
specific heat in the many-component limit$^{19}$, which is largely
equivalent to the Hartree approximation. Up to a non-universal
constant prefactor, the singular part of the specific heat is then
given by
$$
C = 1 - {{\partial \tilde{t}}\over{\partial t}}\ .          \eqno(5.2)
$$

Scaling behaviour of the specific heat now follows directly from that
of $\tilde{t}$.  In the lowest Landau level approximation, we have
$$
C \approx 1 - \tau_{LLL}'(y)\ ,                             \eqno(5.3)
$$
with the scaling variable $y$ and scaling function $\tau_{LLL}$
defined as in (2.12) - (2.13), and this naturally reproduces Bray's
result$^{12}$.  Corresponding to the two-parameter scaling form (4.1)
for $\tilde{t}$, we find
$$
C =1-\hat{\lambda}^{-1}(2eB)^{-\alpha/2\nu}{\cal C}(\theta, \delta)\ ,
                                                            \eqno(5.4)
$$
where $\alpha = 2 - d\nu = -(4-d)/(d-2)$ and
$$
{\cal C}(\theta, \delta) =
{\partial\over{\partial\theta}}\tau_B(\theta , \delta)\ .   \eqno(5.5)
$$
This relation between the scaling functions for the specific heat and
the coherence length is a special feature of the Hartree
approximation, as are the associated relations
$$
\alpha = 1 - 2\nu = -\omega\nu                              \eqno(5.6)
$$
between the specific heat exponent $\alpha$, the coherence length
exponent $\nu$ and the correction to scaling exponent $\omega$.  It
seems plausible, however, that the scaling form (5.4) should be more
generally valid.  That is, we expect that a region should exist in
which both the asymptotic critical behaviour of the specific heat and
the leading corrections are described by a function of the form
$$
C = C_1 - C_2 B^{-\alpha/2\nu}{\cal C}\left({{t+B}\over{B^{1/2\nu}}},\
uB^{\omega/2}\right)\ .                                     \eqno(5.7)
$$
In this expression, $C_1$ and $C_2$ are non-universal amplitudes and
$u$ is the scaling field associated with the leading corrections.
With $t$ and $B$ appropriately scaled (so that, in particular, the
line $H_{c2}(T)$ becomes $t + B = 0$), the scaling function ${\cal C}$
would be universal.

This scaling form has some familiar consequences. In the limit $B
\rightarrow 0$ with $t > 0$, we can use (4.6) to find
$$
C \approx c_1 - c_2t^{-\alpha}\ ,                           \eqno(5.8)
$$
with appropriate (non-universal) constants $c_1$ and $c_2$.  Since
$\alpha$ is negative, the specific heat rises to a cusp at the
critical point $t=0$.  Below $T_c$ in zero field, a real
superconductor is presumably in its Meissner phase, with $\<\phi\>\neq
0$, where our approximations are not valid.  The field dependence at
$T=T_c$ or $t=0$ is given by
$$
C = C_1 -C_2 B^{-\alpha/2\nu}{\cal C}(B^{1-1/2\nu},\ uB^{\omega/2})\ .
                                                            \eqno(5.9)
$$
For small fields, we have the asymptotic power law behaviour
$C \approx c_1 - c_2'B^{-\alpha/2\nu}$.  Corrections to this
asymptotic behaviour now involve both $B^{1-1/2\nu}$ and
$B^{\omega/2}$.  In the Hartree approximation, it happens that
$1-1/2\nu = \omega/2 = (4-d)/2$ (=1/2 in three dimensions).  A real
superconductor, however, is probably characterized by the exponents of
the three-dimensional XY model, for which $\omega/2 \approx 0.4$ and
$1-1/2\nu \approx 0.25$, and the leading correction would be that
involving $B^{1-1/2\nu}$.

Of rather greater interest is the behaviour near the line $t+eB=0$,
corresponding to the mean-field $H_{c2}(T)$ line.  According to our
earlier discussion, one should eventually pass from the critical
region into a region where the lowest Landau level approximation
becomes good.  Within the Hartree approximation, the two-parameter
scaling form (5.4) is exact, and reduces to the LLL scaling form (5.3)
in the limit that $\delta \sim B^{\omega/2}$ is large.  In general,
corrections to the asymptotic, one-parameter critical scaling may have
many contributions beyond those associated with the scaling field $u$
indicated in (5.7), and all of these might be important in the
the region between the critical and LLL regimes.  It is interesting to
speculate, on the other hand, that this may not be so, and that a
crossover to LLL behaviour can be described by the two-parameter
scaling function ${\cal C}(\theta, \delta)$.  This requires that, when
$\delta$ is large,
$$
{\cal C}(\theta, \delta) \approx \delta^{\alpha/\omega\nu}
{\cal C}_{LLL}(\theta\delta^{-(2\nu\phi - 1)/\omega\nu})\ ,\eqno(5.10)
$$
where the argument $\theta\delta^{-(2\nu\phi - 1)/\omega\nu}$
coincides, up to a constant, with the argument $y = (t+eB)/
(\hat{\lambda}f_0eB)^{\phi}$  appearing in (5.3).  We do, of course,
find such behaviour in the Hartree approximation.

Unfortunately, mapping out a two-parameter scaling function
experimentally to test the behaviour suggested in (5.10) would be
extremely difficult.  Theoretically, it is important to note that the
crossover mechanism exemplified by (5.10) is quite different from that
expected at a multicritical point.  In the mean field theory of type
II superconductors, the amplitude of the order parameter in the
Abrikosov vortex lattice vanishes continuously at the line
$H_{c2}(T)$, and this is sometimes described as a second-order phase
transition.  It is possible to speculate that the specific heat, for
example, should exhibit an anomaly along this line, perhaps governed
by a critical exponent $\alpha'$.  In that case, the transition at
$(T, B) = (T_c, 0)$ would be a multicritical point, and one would
expect the one-parameter scaling function to incorporate the anomaly:
$$
{\cal C}(\theta, 0) \sim \vert\theta\vert^{-\alpha'}       \eqno(5.11)
$$
as $\theta \rightarrow 0$.  If such an anomaly exists, it is too weak
to be resolved by any experiment known to us.
\newpage
\noindent{\bf 6.\ Scaling of the conductivity}

Electrical transport properties of a superconductor can be
investigated by using a time-dependent Ginzburg-Landau equation to
describe the dynamics.  A method of calculation is described in detail
by Ullah and Dorsey$^{20}$ who use a Hartree approximation to study
scaling behaviour in the LLL regime.  Using essentially the same
method, we have calculated the conductivities $\sigma_{\parallel}$
and $\sigma_{\perp}$ corresponding to a current parallel or transverse
to the applied magnetic field.  We find that both conductivities can
be written in the two-parameter scaling form
$$
\sigma_a = B^{-(2+z-d)/2}{\cal S}_a(\theta, \delta)\ ,      \eqno(6.1)
$$
with a dynamical exponent $z = 2$, but with different scaling
functions ${\cal S}_{\parallel}$ and ${\cal S}_{\perp}$.  In the limit
of large $\delta$, the LLL scaling properties of the two
conductivities are different, however, and we find
$$
\sigma_{\parallel} \approx {\cal S}_{LLL,\parallel}(y)\ ,   \eqno(6.2)
$$
where ${\cal S}_{LLL, \parallel} = \hbox{const}\times
\tau_{LLL}(y)^{-1/\phi}$, while
$$
\sigma_{\perp} \approx B^{-(z+2 -d)\phi/2}{\cal S}_{LLL, \perp}(y) \ ,
                                                            \eqno(6.3)
$$
with ${\cal S}_{LLL, \perp} = \hbox{const}\times
\tau_{LLL}(y)^{-(z+2-d) /2}$. The results (6.2) and (6.3) agree with
those quoted by Ullah and Dorsey for $d=3$ and $d=2$, except that
their 2-dimensional result for $\sigma_{\parallel}$ is quite different
from (6.2), having the form $\sigma_{\parallel} \approx B^{-1/2} {\cal
S}_{LLL, \parallel}(y)$. We are unable to account for this discrepancy
in detail. No doubt, however, it has to do with the fact that, whereas
we have considered an isotropic $d$-dimensional material, Ullah and
Dorsey deal with a layered 3-dimensional system, and obtain a
two-dimensional limit by taking a large interlayer spacing.

\newpage
\noindent{\bf7.\  Comparison with experiment}

While the Hartree approximation may well provide a useful guide to the
scaling behaviour to be expected in real high-$T_c$ materials, it is
numerically rather inaccurate, having critical exponents $\alpha = -1$
and $\nu =1$ in three dimensions, compared with $\alpha \approx -0.01$
and $\nu \approx 0.67$ for the 3-dimensional XY model, which might be
expected to characterize real materials.  Similarly, we do not expect
the scaling functions to be numerically accurate, and do not present
detailed computations.

Estimates of some gross features of the phase diagram from the Hartree
approximation may, however, have some significance.  To obtain such
estimates, we need to identify the parameters of the model in terms of
measurable quantities.  We use the customary means of identifying the
parameters in (2.1), which is explained, for example, by
Tinkham$^{21}$.  These must be treated with caution, however, since
they are based on the assumption that the entire phase diagram of the
superconductor is described by mean field theory, which is not
actually the case.  In SI units, we find that the constraint equation
(2.8) in 3 dimensions takes the form
$$
\tilde{t} = \xi_0^{-2}\left({T\over{T_c}} -1 \right) + 2\pi
{B\over{\Phi_0}} + 4\pi{{\kappa^2\mu_0k_BT_c}\over{\Phi_0^2}}\left(
{B\over{\Phi_0}}\right)^{1/2}f\left({{\tilde{t}\Phi_0}\over{4\pi B}}
\right)\ ,                                                  \eqno(7.1)
$$
where $\Phi_0 = 2.07\times 10^{-15}$ Wb is the flux quantum, $\mu_0$
is the magnetic permeability, which we take to be that of free space
($\mu_0 = 4\pi \times 10^{-7}$ Hm$^{-1}$) and $\kappa$ is the
Ginzburg-Landau parameter.  The quantity $\xi_0$ is a characteristic
length, of the order of the zero-temperature coherence length.

{}From the first two terms on the right hand side of (7.1), we find the
slope of the $H_{c2}(T)$ line as
$$
{{dB}\over {dT}} = -{{\Phi_0}\over{2\pi\xi_0^2T_c}} =
-3.29\times 10^4(\xi_0^2T_c)^{-1} \ ,                       \eqno(7.2)
$$
where $\xi_0$ is measured in {\AA} and $T_c$ in K.  We can also obtain
an estimate for the field strength, denoted by $B_{LLL}$ in figure 1,
at which the line $H_{c2}(T)$ enters the region of LLL scaling. At
this point, the first two terms of (7.1) cancel, and the argument of
$f$ is equal to a value $\epsilon$, for which $f(\epsilon) \approx
\surd \pi \epsilon^{-1/2}$. We thus find
$$
B_{LLL} = {{\pi(\kappa^2\mu_0k_BT_c)^2}\over{(\Phi_0\epsilon)^3}}
\approx 10^{-13}{{T_c^2\kappa^4}\over {\epsilon^3}}\ ,      \eqno(7.3)
$$
where $B_{LLL}$ is measured in tesla.  There is no clearly defined
value of $\epsilon$ which will guarantee that the LLL approximation is
good.  Certainly, $\epsilon$ must be smaller than the value of
approximately 0.3027 for which $f(\epsilon) = 0$.  Numerically, we
find that $f(z)$ can be reasonably well approximated by a function of
the form $f(z) = f_0(z)z^{-1/2}$ for $0 \le z \le 0.1$, where $f_0(z)$
is an amplitude which varies from $f(0) = \surd \pi$ to $f(0.1) = 1$,
so we surmise that LLL scaling might be good for $z < \epsilon = 0.1$.

Detailed measurements of the fluctuation specific heat of YBaCuO have
recently been reported by Overend{\it et al}$^{\, 22}$. For this
material, $T_c = 92$K, and we may reasonably take $\xi_0 \approx 10$
{\AA} and $\kappa \approx 100$. We then estimate the slope of the
$H_{c2}(T)$ line as $dB/dT \approx -4$ TK$^{-1}$ and, using values of
$\epsilon$ between 0.1 and 0.2, expect that LLL scaling might set in
with an applied field $B_{LLL}$ between 10 and 100 T. In fact, the
measurements of Overend {\it et al} indicate that LLL scaling fails at
all fields up to 8 T. Junod {\it et al}$^{\, 6}$ have recently
reported measurements on the specific heat of YBaCuO up to 20 T. For
fields greater than 2.5 T, they find that the LLL scaling form
collapses their data onto a common curve, but only in a narrow range
of temperature near $T_{c2}(B)$. In the neighbourhood of the peak a
little below this temperature, their data does not scale at all, in
marked contrast to conventional superconductors, where the LLL scaling
region extends well below this peak$^{13, 16}$. For this reason, we do
not think that their data is really consistent with LLL scaling. There
is, however, some indication that the scaling improves at fields of
the order of 20 T. Similarly, earlier specific heat measurements by
Welp {\it et al}$^{\, 17}$ and by Inderhees {\it et al}$^{\, 4}$ are
at best consistent with LLL scaling only in a very narrow range of
temperature.

For a conventional material with, say, $T_c = 10$ K and $\kappa = 10$
our estimate of $B_{LLL}$ would be of the order of 10$^{-4}$ T. Thus,
the schematic phase diagram of figure 1, together with the estimate
(7.3) of $B_{LLL}$ is in reasonable accord with the observation that
LLL scaling appears to work well for conventional superconductors over
a wide range of fields, but seems not to work for high-$T_c$ materials
except, perhaps, in very large fields.

Following earlier authors$^{2,4}$, Overend {\it et al} attempt
to fit their specific heat data to a critical-point scaling expression
of the form
$$
C = C_1 - C_2 B^{-\alpha/2\nu}{\cal C}\left({{T - T_c(0)}\over
{B^{1/2\nu}}}\right) \ ,                                    \eqno(7.4)
$$
using the exponents of the 3-dimensional XY model, and find excellent
agreement for fields up to 8 T, apart from some rounding in very low
fields, which is attributable to finite-size effects. They find,
however, that equally good agreement can be obtained using the scaling
variable $(T - T_{c2}(B))/B^{1/2\nu}$ which appears in (5.7), provided
that the slope of the upper critical field line $-dB_{c2}/dt$ is
greater than about 5 TK$^{-1}$.  This is consistent with the crude
estimate of 4 TK$^{-1}$ obtained above, and with the estimate of about
7 TK$^{-1}$ obtained by Palstra {\it et al}$^{\, 23}$ by extrapolating
the transport entropy of vortex motion to zero. However, it would not
be consistent with the slope of about 1.8 TK$^{-1}$ required by Junod
{\it et al}$^{\, 6}$ to optimize their fits to LLL scaling.

The appearance of the scaling field $T-T_{c2}(B)$, or of $t + eB$ in
(4.2) does not seem to be an artifact of the Hartree approximation
used here. It arises simply from the fact that the Landau eigenvalues
are $k^2 + (2n+1) eB + t$, so that $t$ always appears in the
combination $t+eB$, and this would appear to be true quite generally.
Within the Hartree approximation, we see from (4.5) that the scaling
variable $\theta \sim (T - T_{c2}(B))/B^{1/2\nu}$ is itself a function
of $B/\tilde{t}$ or of $B\xi_B^2$, where $\xi_B$ is the
field-dependent coherence length, rather than the zero-field coherence
length which appears in the gaussian approximation.  The difference
between the two variables $(T - T_c(0))/B^{1/2\nu}$ and $(T -
T_{c2}(B))/B^{1/2\nu}$ is proportional to $B^{(1-1/2\nu)}\approx B^{0.
25}$ if $\nu$ is taken to be the exponent of the 3-dimensional XY
model.  Since this correction varies rather slowly with $B$, the
distinction between the two scaling variables is probably difficult to
discern by optimizing the collapse of data onto a common curve.

\noindent{\bf 8.\ Conclusions}

We have used a simple Hartree approximation to investigate the
critical-point and lowest-Landau-level scaling properties of
high-temperature superconductors.  Our principal conclusions are
summarized in figure 1, which indicates that critical-point scaling is
to be expected in the neighbourhood of the zero-field transition,
while lowest-Landau-level scaling is restricted to a high-field region
near the upper critical field line $H_{c2}(T)$.  In the critical
region, the appropriate scaling variable is $(T-T_{c2}(B))/B^{1/\nu}$
rather than the variable $(T - T_c(0))/B^{1/2\nu}$ which arises in the
gaussian approximation.  While the Hartree approximation does not
yield accurate values for critical exponents or scaling functions, it
suggests that a minimum applied field of between 10 and 100 T is
required to observe lowest-Landau-level scaling in materials such as
YBaCuO.  These conclusions appear to be consistent with current
experimental observations.

I am grateful to Alan Bray, Mark Howson, Michael Moore, Neil Overend
and Myron Salamon for numerous discussions on the issues addressed in
this work.

\newpage
\parindent=0pt
\parskip=0pt
{\bf References}

$^1$D.S. Fisher, M.P.A. Fisher and D.A. Huse, Phys. Rev. B {\bf 43},
130 (1991)

$^2$S.E. Inderhees, M.B. Salamon, N. Goldenfeld, J.P. Rice, B.G.
Pazol and D.M. $^{\ }$Ginzberg, Phys. Rev. Lett. {\bf 60}, 1178 (1988)

$^3$S. Regan, A.J. Lowe and M.A. Howson, J. Phys: Cond. Matt. {\bf
3}, 9245 (1991)

$^4$S.E. Inderhees, M.B. Salamon, J.P. Rice and D.M. Ginzberg, Phys.
Rev. Lett. {\bf 66}, $^{\ }$232 (1991)

$^5$M.B. Salamon, J. Shi, N Overend and M.A. Howson, Phys. Rev. B {\bf
47}, 5520 (1993)

$^6$A. Junod, E. Bonjour, R. Calemczuk, J.Y. Henry, J. Muller,
G. Triscone and J.C. $^{\ }$Vallier, Physica C {\bf 211}, 304 (1993)

$^7$B.I. Halperin, T.C. Lubensky and S-K. Ma, Phys. Rev. Lett. {\bf 32},
292 (1974)

$^8$I.D. Lawrie, Nucl. Phys. B {\bf 200} [FS4], 1 (1982)

$^9$I.D. Lawrie and C. Athorne, J. Phys. A {\bf 16} L587 (1983)

$^{10}$C. Dasgupta and B.I. Halperin, Phys. Rev. Lett. {\bf 47}, 1556
(1981)

$^{11}$R.E. Prange, Phys. Rev. B {\bf 1}, 2349 (1970)

$^{12}$A.J. Bray, Phys. Rev. B {\bf 9}, 4752 (1974)

$^{13}$D.J. Thouless, Phys. Rev. Lett. {\bf 34}, 946 (1975)

$^{\ \ }$S.P. Farrant and C.E. Gough, Phys. Rev. Lett. {\bf 34}, 943
(1975)

$^{14}$G.J. Ruggieri and D.J. Thouless, J. Phys. F {\bf 6}, 2063
(1976)

$^{15}$N.K. Wilkin and M.A. Moore, Phys. Rev. B {\bf 47}, 957 (1993)

$^{16}$N.K. Wilkin and M.A. Moore, Phys. Rev. B {\bf 48}, 3464 (1993)

$^{17}$U. Welp, S. Fleshler, W.K. Kwok, R.A. Klemm, V.M. Vonokur,
J. Downey, B. $^{\ \ }$Veal and G.W. Crabtree, Phys. Rev. Lett. {\bf
67}, 3180 (1991)

$^{18}$S. Hikami and A. Fujita, Phys. Rev. B {\bf 41}, 6379 (1990)

$^{19}$S-K. Ma, in {\it Phase Transitions and Critical Phenomena}, ed.
C. Domb and M.S. $^{\ \ }$Green, Vol. 6 p. 249 (Academic Press:
London, 1976)

$^{20}$S. Ullah and A.T. Dorsey, Phys. Rev. B {\bf 44}, 262 (1991)

$^{21}$M. Tinkham, {\it Introduction to Superconductivity} (McGraw -
Hill:  New York 1975)

$^{22}$N. Overend, M.A. Howson and I.D. Lawrie (to be published)

$^{23}$T.T.M. Palstra, B. Battlogg, L. F. Schneemeyer and J. V.
Waszczak, Phys. Rev. $^{\ \ }$Lett. {\bf 64}, 3090 (1990)
\newpage
{\bf Figure Caption}

Figure 1. Schematic phase diagram of a high-temperature superconductor
in the neighbourhood of its critical point.  Shaded regions are those
in which critical point or lowest-Landau-level scaling should be
observed.  The line $t = -eB$ corresponds to the upper critical field,
while the line $\<\phi\> \sim 0$ is a notional one, below which the
approximation $\<\phi\> \approx 0$ might be expected to fail.  The
curve $\tilde{t}/2eB = \epsilon$ represents a criterion for the
validity of the lowest Landau level approximation explained in the
text.
\end{document}